\PassOptionsToPackage{sort&compress}{natbib}
\documentclass[final,5p,times,twocolumn]{elsarticle}
\usepackage{balance}
\usepackage{amssymb,amsmath,subcaption}
\usepackage[colorlinks=true,linkcolor=blue,citecolor=blue,urlcolor=blue]{hyperref}
\usepackage{lipsum,bm,slashed,graphicx,float,caption,cancel}
\usepackage{xcolor,color,soul,ulem}

\journal{Physics Letters B}

\begin{document}
	
	\begin{frontmatter}

	\title{Non-perturbative heavy quark diffusion coefficients in arbitrarily magnetized quark-gluon plasma}
		
	\author[1]{Debarshi Dey\corref{cor1}}
	\ead{ddey@sjtu.edu.cn}
	
	\author[2,3]{Aritra Bandyopadhyay\corref{cor1}}
	\ead{aritra.bandyopadhyay@e-uvt.ro}

    \author[1]{Yifeng Sun\corref{cor1}}
	\ead{sunyfphy@sjtu.edu.cn}
	
	\author[4]{Santosh K. Das\corref{cor1}}
	\ead{santosh@iitgoa.ac.in}

	\affiliation[1]{organization={State Key Laboratory of Dark Matter Physics, Shanghai Key Laboratory for Particle Physics and Cosmology,
Key Laboratory for Particle Astrophysics and Cosmology (MOE),
School of Physics \& Astronomy, Shanghai Jiao Tong University},
		city={Shanghai},
		postcode={200240},
		country={China}}

    \affiliation[2]{organization={Institute of Theoretical Physics, University of Wrocław},
		addressline={plac Maksa Borna 9},
		city={Wrocław},
		postcode={50204},
		country={Poland}}

	\affiliation[3]{organization={Department of Physics, West University of Timişoara},
		addressline={Bd. Vasile Pârvan 4},
		city={Timişoara},
		postcode={300223},
		country={Romania}}

	\affiliation[4]{organization={School of Physical Sciences, Indian Institute of Technology Goa},
		city={Ponda},
		postcode={403401},
		state={Goa},
		country={India}}
	
			\cortext[cor1]{Corresponding author.}

		\begin{abstract}
		Heavy quark (HQ) momentum ($\kappa$) and spatial diffusion ($D_s$) coefficients are computed in a non-perturbative thermal QCD medium in the presence of a background magnetic field of arbitrary strength. Both perturbative and non-perturbative effects are incorporated via the in-medium HQ potential, obtained from the resummed gluon propagator. We find that the momentum diffusion coefficients become anisotropic even in the static heavy quark limit, with the magnetic field direction defining the axis of anisotropy. This anisotropy originates from restrictions on longitudinal momentum diffusion in the gluon spectral function, and naturally leads to two spatial diffusion coefficients ($D_s^L$, $D_s^T$). Non-perturbative effects are found to be dominant at low temperatures. These results provide a more consistent input for Langevin based calculations of the heavy quark directed flow at RHIC and LHC energies.
		\end{abstract}
		
		
		

	\end{frontmatter}
	
	
	
	
	\section{Introduction}
	\label{introduction}
	
Heavy quarks (HQ) such as charm ($c$) and bottom ($b$) have long been considered valuable probes of the matter created in heavy-ion collisions~\cite{Rapp:2018qla,Das:2406.2024}. 
Being much heavier than the light quarks confers on them certain advantages, one of which is that they are produced in hard scattering processes in the initial stages of heavy-ion collisions, after which their numbers remain essentially fixed, owing to negligible thermal production in the medium. Furthermore,  their thermalization times are comparable to or larger than the QGP lifetime, and so, heavy hadrons retain a memory of their interactions, and their spectra can serve as markers of the interaction strength of the medium~\cite{Dong:2019unq,QGP4,He:PPNP130'2023,Cao:2018ews}. 

In the recent past, the interplay between HQ dynamics and the intense initial electromagnetic fields generated in off-central heavy-ion collisions has generated a lot of interest in the heavy-ion research community~\cite{Skokov:2009qp, Voronyuk:2011jd,Rybicki:PRC872013,Gursoy:PRC89_2014, DAS:PLB768'2017,Sun:2023adv,Chatterjee:PLB798'2019}. In addition to modifying HQ dynamics in the medium, these fields can also be probed using heavy quarks. In particular, the directed flow of heavy flavor, $v_1$, and its rapidity dependence $y$, was proposed as a sensitive probe of the initial electromagnetic fields in Ref.~\cite{DAS:PLB768'2017}. HQ directed flow is predicted in Ref.~\cite{DAS:PLB768'2017} to be roughly an order of magnitude larger than that of light hadrons. However, to isolate other contributions to directed flow, the electromagnetic-field-induced splitting between positively and negatively charged quarks has been proposed as a sensitive observable. In particular, the difference in directed flow between charm and anti-charm quarks, studied via $D^0$ and $\overline{D}^0$ mesons, provides a novel means to probe and quantify the initial electromagnetic field. Subsequently, experimental measurements were reported by the STAR~\cite{STAR:2019clv} and ALICE~\cite{ALICE:2019sgg} collaborations at RHIC and the LHC, respectively. Both collaborations have measured a finite $v_1$ for the $D$-meson. Several theoretical attempts have been made over the years~\cite{DAS:PLB768'2017,Chatterjee:PLB798'2019, Oliva:2020mfr,Sun:2020wkg,Oliva:2020doe,Dubla:2020bdz,Beraudo:2021ont,Jiang:2022uoe,Shen:2025unr,Das:2025yxy,Panda:2026kko}; however, the slope of the $D$-meson directed flow splitting is still not fully understood.

For instance, while the STAR collaboration reported negative $v_1$ slopes ($dv_1/dy$) for both   $D_0$ and $\overline{D}_0$ mesons, the ALICE collaboration reported a positive slope for $D_0$ and a negative slope for $\overline{D}_0$. They also reported a $v_1$ of $D_0$ and $\overline{D}_0$ that was one order of magnitude larger than the theoretical predictions~\cite{DAS:PLB768'2017,Chatterjee:PRL120'2018,Chatterjee:PLB798'2019}.  
Further, the slope of $\Delta v_1$\footnote{$\Delta v_1=v_1\,\,\text{of}\,\,D_0-v_1\,\,of\,\,\overline{D}_0$. In general, it is the subtraction of antiparticle $v_1$ from particle $v_1$.}, $i.e.$ $d\Delta v_1/dy$ was reported to be positive in the ALICE study ($2.7\sigma$ confidence), whereas it was reported to be negative in the STAR  results, as well as in the theory result of ~\cite{DAS:PLB768'2017}. Most models have obtained a finite but negative slope for the D-meson directed flow splitting at both RHIC and LHC energies. 
The slope of $\Delta v_1$ at RHIC and LHC has opposite signs, and both the slope and magnitude of $\Delta v_1$ at LHC energies are still not well understood. 
This indicates the need for a more detailed theoretical investigation of $v_1$ for HQ, which is typically computed using the Langevin equation with an external Lorentz force, where the HQ drag and diffusion coefficients serve as input parameters. However, for a more rigorous treatment, these transport coefficients themselves must be evaluated in the presence of a magnetic field. This provides the phenomenological motivation to compute the HQ drag and diffusion coefficients in an arbitrarily magnetized thermal QCD medium using quantum field theory techniques. 

On the theoretical front, there have been studies of HQ dynamics in a magnetized thermal QCD medium~\cite{Fukushima:PRD93'2016,Kurian:2019nna,Singh:arxiv'2020,Singh:JHEP5'2020,Kurian:2020kct,Bandyopadhyay:PRD105'2022,Dey:PRD109'2024,Satapathy:PRC109'2024,Das:2024wht,Jamal_2024}. However, most of these studies are carried out in the limit of either strong ($eB\gg T^2 $) or weak  ($eB\ll T^2$). In this work, building on the results of ~\cite{Bandyopadhyay:2023hiv}, we use the Landau quantization framework which renders the study valid for an arbitrary strength of the magnetic field. The key result is that the momentum diffusion coefficients ($\kappa$) in the zero momentum limit of the HQ ($\bm{p}\to 0$) are no longer isotropic. We elaborate on the origin of this anisotropy below. 

In general, the direction of the HQ velocity ($\bm{v}$) provides an anisotropy direction to the HQ momentum diffusion, $i.e.,$ the magnitudes of $\kappa$ along and transverse to $v$ are different. The addition of a background magnetic field adds another direction of anisotropy-the direction of the magnetic field. The interplay between these two anisotropy directions determines the structure of $\kappa$~\cite{Bandyopadhyay:PRD105'2022}. Thus, in the limit $\bm{v}\to 0$, and in the absence of any external fields,  $\kappa$ is isotropic~\cite{Moore:PRC71'2005}. Even in the presence of a magnetic field, this isotropy is preserved in the static HQ limit if the field strength is weak ($eB\ll T^2$)~\cite{Dey:PRD109'2024,Dey:PRD112'2025}. In the present work, however, we allow the magnetic field strength to vary from arbitrarily small to arbitrarily large values and find that $\kappa$ becomes anisotropic even in the static HQ limit. This indicates that, within our framework, the magnetic field alone is sufficient to induce anisotropic momentum diffusion, irrespective of its strength.

In this work, the in-medium HQ potential $V(q)$ in a magnetized medium is first derived~\cite{Indrani:EPJC83_2023,Hasan:PRD102_2020}. This is done using the gluon propagator, the temporal component of which is parametrised as~\cite{Eugenio:JHEP1'2006,Megias:PRD75'2007,Riek:PRC82'2010}
\begin{equation}\label{np_gp}
	D_{00}(Q)=D^{\text{p}}_{00}(Q)+D^{\text{str}}_{00}(Q),
\end{equation}
where, $Q$ is the 4-momentum of the gluon. Here, the superscripts p and  str refer to perturbative and  string, respectively. Next, the HQ self-energy $\Sigma(P)$ is computed, which is given by
\begin{equation}
	\Sigma(P)=i g^2 \int \frac{d^4 Q}{(2 \pi)^4} D^{\mu \nu}(Q) \gamma_\mu S_m(P-Q) \gamma_\nu,\label{self_energy}
\end{equation}
with $g$ as the strong coupling constant, and $P$, the 4-momentum of the HQ. Here, $D^{\mu \nu}(Q)$ is the resummed gluon propagator, and $S_m(P-Q)$ is the HQ propagator, both in the presence of magnetic field. $V(q)$ is used as a proxy for $D^{\mu\nu}$ in the static limit of the HQ to evaluate $\Sigma$, as is detailed in Section~\ref{transport}. From $\Sigma(P)$,  the HQ scattering rate is evaluated using Weldon's formula:~\cite{Weldon:PRD28'1983}
\begin{equation}\label{gamma}
	\Gamma(E , \bm{v})=-\frac{1}{2 E} [1-n_F(E)] \operatorname{Tr}\left[(\slashed{P}+M_Q) \operatorname{Im} \Sigma\left(p_0+i \epsilon, \bm{p}\right)\right],
\end{equation}
Using $\Gamma$, one then evaluates the momentum diffusion coefficients via
\begin{align}\label{kappa}
	\kappa_{i}=\int d^3 q \frac{d \Gamma(E,v)}{d^3 q} q_{i}^2.
\end{align} 
The paper is organized as follows:  In Section~\ref{potential}, the evaluation of the HQ in-medium potential is outlined. This is followed by the calculation of the HQ transport coefficients in Section~\ref{transport}. Finally, the summary and conclusions are presented in Section~\ref{summary}.
	
\section{In-medium heavy quark potential}\label{potential}
The in-medium potential $V(q)$ is obtained by appropriately weighting the vacuum potential  $V_0(q)$ with the medium permittivity $\epsilon(q)$: 
\begin{equation}\label{pot}
	V(q)=V_0^{\text{p}}(q)\, \frac{1}{\epsilon(q)}
\end{equation}
The permittivity, in turn is evaluated from the gauge boson propagator, and encodes the information about the medium~\cite{ThakurPhysRevD.97.096011,WeldonPhysRevD.26.1394}  
\begin{equation}
	\frac{1}{\epsilon(q)}=-q^2D_{00}(q_0=0,q)\label{eps}
\end{equation}
A non-perturbative ansatz is added to the usual gluon propagator with the constraint that it reproduces the well-known confining term of the Cornell-type vacuum HQ potential in position space,\footnote{$V_0(r)=-\frac{4}{3}\frac{\alpha_s}{r}+\sigma r\equiv V_0^{\text{p}}(r)+V_0^{\text{str}}(r)$. Here, $\alpha_s = g^2/(4\pi)$, and $\sigma r$ is the confining term with $\sigma$ being the string tension.} \textit{i.e.}, one ought to have $D_{\mu\nu}(Q)=D_{\mu\nu}^{\text{p}}(Q)+D_{\mu\nu}^{\text{str}}(Q)$. For static heavy quarks, however, only $D^{00}$ contributes in Eq.~\eqref{self_energy}, and it suffices to parametrize only the temporal component of the propagator via Eq.~\eqref{np_gp}. One now needs to compute both $D^{\text{p}}_{00}(Q)$ and $D^{\text{str}}_{00}(Q)$ in the presence of an arbitrary magnetic field. 

For the perturbative part, we write the usual Schwinger-Dyson equation (SDE) relating the full (resummed) propagator, the bare propagator and the one-loop self energy:
\begin{equation}
	D_{\mu\nu}(q_0,q)=\left[Q^2g_{\mu\nu}-\Pi_{\mu\nu}(q_0,q)\right]^{-1}, 
\end{equation}
where, $\Pi_{\mu\nu}$ can be expanded generally in a rank-2 tensor basis as~\cite{Karmakar:EPJC79'2019}
\begin{multline}
	\Pi^{\mu\nu}(q_0,q)=d_1(q_0,q)\Delta_1^{\mu\nu}+d_2(q_0,q)\Delta_2^{\mu\nu}+\\d_3(q_0,q)\Delta_3^{\mu\nu}+d_4(q_0,q)\Delta_4^{\mu\nu},\label{ff}
\end{multline} 
with $d_i$ being the Lorentz invariant form factors. As mentioned earlier, for static HQ, only the temporal (00) component contributes. This leads to\footnote{See~\cite{Karmakar:EPJC79'2019} for the structure of the basis tensors. For the $00$ component, only $\Delta_1^{\mu\nu}$ survives.  $\bar{u}^{\mu}=u^{\mu}-\frac{q^0Q^{\mu}}{Q^2};$ $u^{\mu}$ is the velocity of the heat bath, which, in the local rest frame reads $(1,0,0,0)$.} 
\begin{equation}
	D_{00}^{\text{p}}=\left[Q^2-d_1\bar{u}^2\right]^{-1}.
\end{equation}
The self energy, and hence, the form factors are originally computed in Euclidean space and then analytically continued to real energies via $q_0\to \omega+i\epsilon$, rendering them complex. The analytical continuation can be recognized as leading to the retarded propagator, as should be the case, since Eq.~\eqref{gamma} uses the retarded self energy. This is understandable because the retarded self energy is the linear response function of the field to an external source, and that is what we are interested in, for HQ scattering. Thus we can express $d_1$ as $d_1=a+ib$. On rationalization, one obtains
\begin{equation}
	\text{Re}D^{\text{p}}_{00}=\frac{Q^2+a}{(Q^2+a)^2+b^2}\,,\quad  \text{Im}D^{\text{p}}_{00}=\frac{-b}{(Q^2+a)^2+b^2}.
\end{equation}
Recall that all expressions need to be in the static limit. From Eqs. [(E11), (E12)] of~\cite{Bandyopadhyay:2023hiv}, we see that 
\begin{align}
	a(\omega=0)&=m_D'^2=(m_D^g)^2+\sum_f \delta m_{D,f}^2\\
	\frac{b}{\omega}(\omega\to0)&=(m_D^g)^2\frac{\pi}{2q}+\frac{\pi}{2}\delta(q_z)\sum_f\delta m_{D,f}^2,\label{img_p1}
\end{align}
where,
\begin{align}
(m_D^g)^2&=N_cg^2T^2/3 \, ,\\
\delta m_{D,f}^2&=\frac{g^2|q_fB|}{4\pi^2T}\sum_{l=0}^{\infty}(2-\delta_{l,0})\int dk_z n_F(1-n_F) \, ,
\end{align}
$N_c$ being the number of colors. This yields
\begin{equation}\label{red00p}
	\text{Re}D^{\text{p}}_{00}(\omega=0)=\frac{1}{q^2+m_D'^2}
\end{equation}
For the imaginary part, we have
\begin{align}
	\frac{\text{Im}D^{\text{p}}_{00}}{\omega}(\omega\to 0)&=\lim_{\omega\to 0}\frac{1}{\omega}\frac{(-b)}{(Q^2+a)^2+b^2}\nonumber\\[0.3em]
	&=-\pi \frac{(m_D^g)^2/q+\delta(q_z)\sum_f\delta m_{D,f}^2}{2(q^2+m_D'^2)^2}.\label{imd00p}
\end{align}
We note that the $\omega$ in the denominator of Eq.~\eqref{imd00p} will be provided by the small $\omega$ expansion of the factor $1+n_B(\omega)$\footnote{$[1+n_B(\omega)](\omega\to 0)=T/\omega+1/2-\mathcal{O}(\omega/T)+\mathcal{O}(\omega/T)^2+\cdots\approx T/\omega$} in Eq.~\eqref{sfpot}.  As a minimal extension of the perturbative Schwinger-Dyson equation, one parametrizes the string propagator as~\cite{Guo:PRD100'2019}
\begin{equation}\label{imd1}
	D_{00}^{\text{str}}(Q)=m_G^2\left[Q^2-\Pi_{00}(Q)\right]^{-2},
\end{equation}
where $m_G^2$ is a dimensional constant (of dimension $M^2$) related to the string tension $\sigma$. For the potential to reduce to the well known vacuum result, $\sigma$ should be related to $m_G^2$ as $\sigma=2/3\alpha_s m_G^2$. For our calculations we have used the data of the $T$ dependent spatial string tension presented in \cite{BalaPRL135:2025}. Note that the self-energy in Eq.~\eqref{imd1} is still the perturbative self energy, and so, the form factors are the same as in Eq.~\eqref{ff}. Thus, we have
\begin{equation}
	D_{00}^{\text{str}}=m_G^2\left[Q^2-d_1\bar{u}^2\right]^{-2}.
\end{equation}
Following the same procedure as earlier, we arrive at
\begin{align}
	\text{Re}D^{\text{str}}_{00}(\omega= 0)&=\frac{m_G^2}{(q^2+m_D'^2)^2}\label{red00np}\\ \frac{\text{Im}D^{\text{str}}_{00}}{\omega}(\omega\to 0)&=-\frac{\pi m_G^2\left[(m_D^g)^2/q+\delta(q_z)\sum_f\delta m_{D,f}^2\right]}{(q^2+m_D'^2)^3}.\label{imd00np}
\end{align}
Using Eqs.\eqref{red00p},\eqref{imd00p},\eqref{red00np}, and \eqref{imd00np} in Eq.~\eqref{eps}, one obtains the real and imaginary parts of the perturbative ($\epsilon^Y$) and non-perturbative ($\epsilon^s$) components\footnote{The superscripts $Y$ and $\text{str}$ stand for Yukawa and string, respectively.} of $\epsilon(q)$, which then leads to the in-medium potential via Eq.~\eqref{pot}. Thus, the complete expression of the in-medium heavy quark potential turns out to be:
\begin{align}
	V^Y(q)=-\frac{4}{3}g^2\left[\frac{1}{q^2+m_D'^2}-i\pi \frac{(m_D^g)^2/q+\delta(q_z)\sum_f  \delta m_{D,f}^2}{2(q^2+m_D'^2)^2}\right]\label{vy}
\end{align}
\begin{multline}
	V^s(q)=-\frac{4}{3}g^2\Bigg[\frac{m_G^2}{(q^2+m_D'^2)^2}-\\
	i\pi m_G^2\frac{(m_D^g)^2/q+\delta(q_z)\sum_f  \delta m_{D,f}^2}{q(q^2+m_D'^2)^3}\Bigg]\label{vS},
\end{multline}
\begin{equation}\label{pot_tot}
	\text{with},\quad   V(q)=V^Y(q)+V^{\text{str}}(q).
\end{equation}
	

\section{HQ momentum and spatial diffusion coefficients}\label{transport}

The HQ potential in Eq.~\eqref{pot_tot} serves as a  proxy for the gluon propagator in Eq.~\eqref{self_energy}. Because of the form of the potential, the self-energy in Eq.~\eqref{self_energy} also has a perturbative (Yukawa) and a non-perturbative (string) part. For the Yukawa part, the potential-propagator interchange is achieved via the prescription $g^2D^{\mu\nu}\rightarrow V^Y(q)$, so that the perturbative self-energy reads
\begin{equation}\label{sigmaY}
	\Sigma^Y(P)=i  \int \frac{d^4 Q}{(2 \pi)^4} V^Y(q) \gamma_\mu S(P-Q) \gamma^\mu.
\end{equation}
For the string part, a scalar interaction vertex is considered in addition, and so, the Lorentz gamma matrices do not feature:
\begin{equation}\label{sigmaS}
	\Sigma^{\text{str}}(P)=i  \int \frac{d^4 Q}{(2 \pi)^4} V^{\text{str}}(q)S(P-Q),
\end{equation}
where,
\begin{equation}
	S_m(P-Q\equiv K) = e^{-\frac{k_{\perp}^2}{|q_f B|}} \sum_{l=0}^{\infty} \frac{(-1)^l D_l(q_f B, K)}{K_{\parallel}^2 - M_Q^2 - 2l q_f B},
\end{equation}
where, $l=0,1,2,\cdots$ denotes the Landau levels, $K_{\parallel}^{\mu}=(k^0,0,0,k_z)$, $q_f$ is the electric charge of the fermion, and
\begin{equation}
	\begin{aligned}
		D_l(q_f B, K) = {} & (\slashed{K}_{\parallel} + M) ((1 - i\gamma^1 \gamma^2) L_l(\xi_k^{\perp}) \\
		& - (1 + i\gamma^1 \gamma^2) L_{l-1}(\xi_k^{\perp})) - 4\slashed{k}_{\perp} L_{l-1}^1(\xi_k^{\perp}),
	\end{aligned}
\end{equation}
with $\xi^{\perp}_k=\frac{2k_{\perp}^2}{q_fB}$, $K_{\perp}^{\mu}=(0,k_x,k_y,0)$ and $L_l^{\alpha}$ the generalised Laguerre polynomial defined as
\begin{equation}
	(1 - z)^{-(\alpha+1)} \exp \left( \frac{z \xi_k^{\perp}}{z - 1} \right) = \sum_{l=0}^{\infty} L_l^{\alpha}(\xi_k^{\perp}) z^l.
\end{equation}
Next, we use the spectral representation of the potential, given by
\begin{align}\label{sfpot}
	V^{Y/{\text{str}}}(q)&=-\int_{0}^{\beta}d\tau\,e^{q_0\tau}\int_{-\infty}^{\infty}d\omega\,\rho^{Y/{\text{str}}}(q)\left[1+n_B(\omega)\right]e^{-\omega\tau},
\end{align}
with
\begin{equation}
	\rho^{Y/s}(q)=-\frac{\text{Im}\,V^{Y/{\text{str}}}(q)}{\pi},
\end{equation}
being the spectral function corresponding to Yukawa/String term. This is done primarily to make the Matsubara frequency sum over $q_0$ convenient. 
Using these results and after some algebra, Eq.~\eqref{gamma} becomes:
\begin{multline}
	\Gamma^{Y/\text{str}}(E,\bm{v})=\frac{T\pi }{4E^2}\int\frac{d^3 q}{(2 \pi)^3}  \int_{-\infty}^{+\infty} d \omega\rho^{Y/\text{str}}(q)\\
	A^{Y/\text{str}}(\omega)\,\delta(\omega-\bm{v}\cdot\bm{q}),\label{SRfinal}
\end{multline}
with $A^Y=8(M_Q^2-\bm{p}\cdot \bm{q})$, $A^{\text{str}}=4(2M_Q^2+\bm{p}\cdot \bm{q})$ being the contributions due to Dirac traces in Eq.~\eqref{sigmaY} and Eq.~\eqref{sigmaS}, respectively.
For convenience, we split the spectral function as a sum of two terms :
\begin{align}
	\rho^Y_1&=-\frac{4}{3}g^2\frac{  (m_D^g)^2}{2q(q^2+m_D'^2)^2},\quad \rho^Y_2=-\frac{4}{3}g^2\frac{ \delta(q_z)\sum_f\delta m_{D,f}^2}{2(q^2+m_D'^2)^2}\\
	\rho^{\text{str}}_1&=-\frac{4}{3}g^2\frac{ m_G^2 (m_D^g)^2}{q(q^2+m_D'^2)^3},\quad \rho^{\text{str}}_2=-\frac{4}{3}g^2\frac{m_G^2\delta(q_z)\sum_f\delta m_{D,f}^2}{(q^2+m_D'^2)^3}\label{rho_s},
\end{align}
where, $\rho^{Y/s}=\rho_1^{Y/\text{str}}+\rho_2^{Y/\text{str}}$, noting that $\rho_1^{Y/\text{str}}$ and $\rho_2^{Y/\text{str}}$ have distinct angular structures which, as we shall see, is the origin of the anisotropy between $\kappa_L$ and $\kappa_T$.   Since we consider static heavy quarks, the external magnetic field ($\bm{B}=B\,\hat{z}$) solely gives rise to the anisotropy in the system, and so, we define momentum diffusion coefficients along and transverse to the direction of the magnetic field.
\begin{equation}
	\kappa_{L}=\int d^3 q \frac{d \Gamma(E,v)}{d^3 q} q_{z}^2\,,\quad 
	\kappa_{T}=\frac{1}{2}\int d^3 q \frac{d \Gamma(E,v)}{d^3 q} \bm{q_{\perp}}^2,
\end{equation}
where, $\bm{q_{\perp}^2}=q_x^2+q_y^2$. 
In the static limit, $v=0$, and we have $A^Y=A^s\equiv A=8M_Q^2$, which leads to
\begin{equation}
	\kappa_L=\frac{AT}{16E^2\pi}\int_0^{\infty}dq\int_0^{\pi}d\theta\,q^4\sin\theta\cos^2\theta\{\rho_1(q)+\rho_2(q,\theta)\}  
\end{equation}
$\rho_2(q,\theta)$ contains $\delta(q_z)=\delta(q\cos\theta)$, because of which, the $\theta$ integration of the corresponding integral is
\begin{equation}
	I_\theta^L= \int_0^\pi d\theta\,\sin\theta\cos^2\theta\,\delta(q\cos\theta)=0.
\end{equation}
Thus, $\rho_2^{Y/\text{str}}$ does not contribute to $\kappa_L^{Y/\text{str}}$, while it does contribute to $\kappa_T^{Y/\text{str}}$. This is the source of anisotropy between $\kappa_L$ and $\kappa_T$. Thus, we have
\begin{equation}
	\kappa_L^Y=-\frac{g^2AT(m_D^g)^2}{36\pi E^2}\int_0^\infty dq\,\frac{q^3}{(q^2+m_D'^2)^2}
\end{equation}
For $\kappa_T^Y$, the $\theta$ integration term is
\begin{equation}
	I_\theta^T=\int_0^\pi d\theta \sin^3\theta\,\delta(q\cos\theta)=\frac{1}{q},
\end{equation}
and so, we have
\begin{equation}
	\kappa_T^Y=\kappa_L^Y-\frac{g^2AT\sum_f\delta m_{D,f}^2}{48\pi E^2}\int\limits_0^\infty dq\frac{q^3}{(q^2+m_D'^2)^2}.
\end{equation}
We see that the same $q$ integration appears in both the cases:
\begin{equation}
	I_q^Y=\int\limits_0^\infty dq \frac{q^3}{(q^2+m_D'^2)^2}.
\end{equation}
The integral, and hence, the $\kappa$ are UV divergent. This is well known, and arises because our calculations are confined to
the region of soft gauge boson momentum transfer\footnote{ One defines an intermediate momentum scale $q*$, which serves as the lower limit of integration for hard contribution, and as the upper limit for soft contribution. On adding the contributions, the $q*$ dependence cancels out in the final result~\cite{Braaten:PRD44'1991}. A similar exercise at $B\neq 0$ is left for a future work.}~\cite{Braaten:PRD44'1991}. Accordingly, a UV cut-off $q_{max}=3.1Tg^{1/3}$ \cite{Beraudo:NPA831'2009} is applied while evaluating the integrals, so that we have
\begin{equation}
 I_q^Y = \frac{1}{2} \left[\ln{|q^2+m_D'^2|}+ \frac{m_D'^2}{q^2+m_D'^2} \right]_0^{q_{max}} .  
\end{equation}
For the string part, the relevant integral can be seen from Eq.~\eqref{rho_s}. It is 
\begin{align}
	I_q^{\text{str}}=\int\limits_0^\infty dq \frac{q^3}{(q^2+m_D'^2)^3}=\frac{1}{4m_D'^2}.
\end{align}
Because of two extra powers of $q$ in the denominator, this integral is UV finite. Thus, the expression for the string parts are
\begin{align}
		\kappa_L^{\text{str}}&=-\frac{g^2AT(m_D^g)^2m_G^2}{72\pi E^2m_D'^2}\\
		\kappa_T^{\text{str}}&=\kappa_L^{\text{str}}-\frac{g^2ATm_G^2\sum_f\delta m_{D,f}^2}{96\pi E^2m_D'^2}.
	\end{align}
Both the Yukawa and the string parts of $\kappa$ are anisotropic, as can be clearly seen in the analytic expressions.

\begin{figure*}[t] 
	\centering
	\begin{subfigure}{0.46\textwidth}
		\centering
		\includegraphics[width=\textwidth,height=6cm]{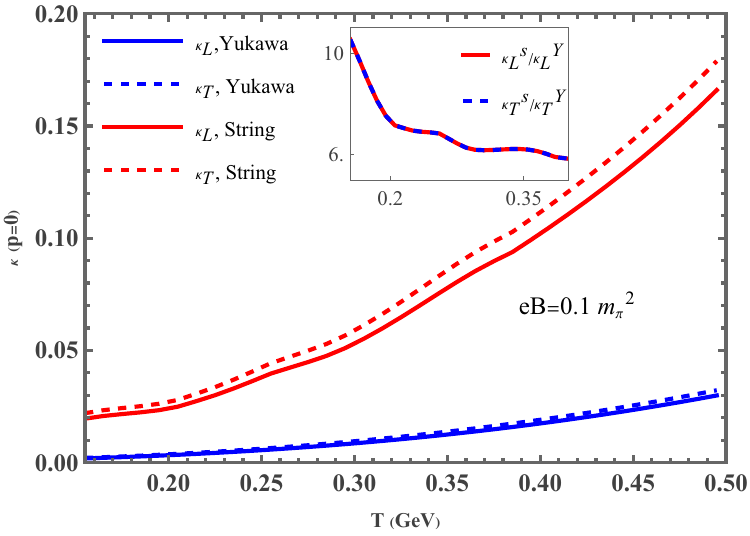}
		\caption{Yukawa and string components of the momentum diffusion coefficient ($\kappa$) in the static limit as a function of temperature. The inset shows the ratio of the string to Yukawa contributions of both the longitudinal and transverse $\kappa$.}
		\label{ratio}
	\end{subfigure}
	\hfill 
	\begin{subfigure}{0.46\textwidth}
		\centering
		\includegraphics[width=\textwidth,height=6cm]{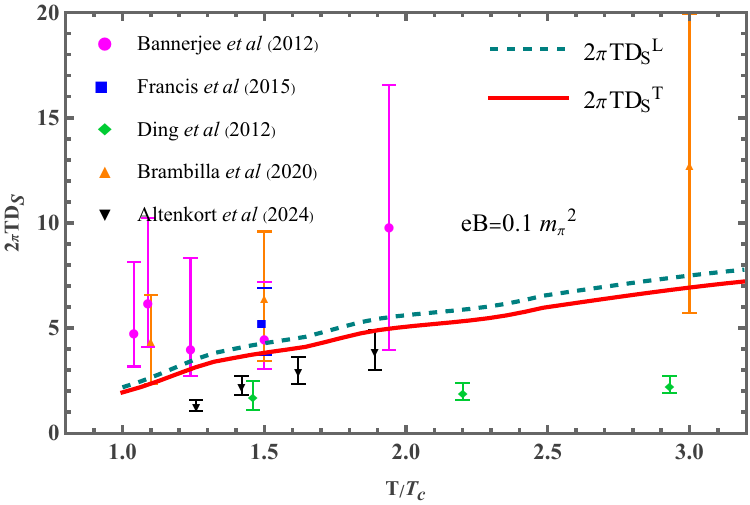} 
		\caption{Scaled spatial diffusion coefficient as a function of $T$ at $eB=0.1\,m_{\pi}^2$. Several lattice results are also shown for comparison~\cite{Banerjee:PRD85'2012,Francis:PRD92'2015,Ding:PRD86'2012,Brambilla:PRD102'2020,Altenkort:PRL132'2024}. The Altenkort result is HQ mass dependent; their charm quark results are shown in the figure.}
		\label{static limit}
	\end{subfigure}
	\caption{}
\end{figure*}

The spatial diffusion coefficient is defined as~\cite{He:PPNP130'2023}
\begin{equation}
	D_s=\frac{2T^2}{\kappa(p=0)},
\end{equation}
with $\kappa=\kappa^Y+\kappa^s$. In the absence of a magnetic field, $\kappa$ is isotropic 
in the $p\to 0$ limit, yielding a single spatial diffusion 
coefficient. In the present case, however, the anisotropy 
persists even in the $p\to 0$ limit, necessitating two 
distinct spatial diffusion coefficients $D_s^L$ and $D_s^T$, 
corresponding to $\kappa_L$ and $\kappa_T$, respectively. From Fig.~\eqref{ratio} with $eB=0.1\,m_\pi^2$, it can be seen that the anisotropy is comparatively more pronounced for the string part. Interestingly however, the inset graph reveals that the ratio of the string to Yukawa contribution is indeed isotropic. We thus have $\kappa_L^Y\neq \kappa_T^Y\neq \kappa_L^{\text{str}}\neq\kappa_T^{\text{str}}$, but $\frac{\kappa_L^{\text{str}}}{\kappa^Y_L}=\frac{\kappa_T^{\text{str}}}{\kappa^Y_T}$. Further, the inset also shows that the relative contribution of the non-perturbative component is much larger at low $T$. Fig.~\eqref{static limit} shows the scaled spatial diffusion coefficients $2\pi TD_s^L$ and $2\pi TD_s^T$ as functions of $T/T_c$ at $eB=0.1\,m_\pi^2$. The two coefficients are clearly separated, with the longitudinal coefficient being slightly larger than the transverse one throughout the temperature range shown. It should also be mentioned that the magnitudes of the coefficients have a very mild dependence on the strength of the external magnetic field in the weak magnetic field regime. Our results show broad agreement with the available lattice data, most notably in the range $T\lesssim 2T_c$. 

\section{Summary and Conclusion}\label{summary}

In this work, we have evaluated the momentum and spatial diffusion coefficients of static charm quark in a non-perturbative thermal QCD medium in the presence of a background magnetic field of arbitrary strength. To that end, we first calculate the HQ in-medium potential ($V$) from the non-perturbative ansatz of the gauge boson propagator. Thereafter, the potential is used as a proxy for the gauge boson propagator in the evaluation of the HQ self-energy ($\Sigma$). From the self-energy, we then calculate the HQ longitudinal and transverse momentum diffusion coefficients, and thereafter, the spatial diffusion coefficients.

We find that in the presence of an arbitrary magnetic field, the momentum diffusion of even static heavy quarks can be anisotropic, with the magnetic field direction defining the direction of anisotropy. The root of this anisotropy lies in the delta function $\delta(q_z)$ in the imaginary part of the form factor $d_1$ in the one loop gluon self-energy, effectively prohibiting longitudinal momentum diffusion from a part of the spectral function, while on the other hand, the entire spectral function contributes to the transverse momentum diffusion. Consequently, the anisotropy shows up in the spatial diffusion coefficient ($D_s$) as well, which leads us to define $D_s^L$ and $D_s^T$ for longitudinal and transverse spatial diffusion coefficients. However, we find the ratio of the string to Yukawa contribution of $\kappa$ to be isotropic, as can be seen from the inset Fig.~\eqref{ratio}. The scaled spatial diffusion coefficients $2\pi TD_s^L$ and $2\pi TD_s^T$ show broad qualitative agreement with available lattice QCD results, noting that the lattice results are obtained at zero magnetic field. Lastly, although the anisotropy structure of the coefficients is rather interesting, the absolute magnitudes of the coefficients depend very mildly on the magnetic field strength in the weak magnetic field regime. These results provide a more consistent input for the Langevin-based calculations of heavy quark directed flow $v_1$ at RHIC and LHC energies, where the interplay between the magnetic field and the non-perturbative medium effects is expected to play an important role.

	\section*{Acknowledgements}
This work was supported by the National Natural Science Foundation of China (NSFC) under Grant
Nos. 12422508, 12375124, and the Science and Technology Commission of Shanghai Municipality under Grant No. 23JC1402700. A.~B. acknowledges support from the ULAM fellowship program of the Polish National Agency for Academic Exchange (NAWA), No.~BNI/ULM/2024/1/00193 and EU’s NextGenerationEU instrument through the National Recovery and Resilience Plan of Romania - Pillar III-C9-I8, managed by the Ministry of Research, Innovation and Digitization, within the project entitled ``Facets of Rotating Quark-Gluon Plasma'' (FORQ), contract no.~760079/23.05.2023 code CF 103/15.11.2022. Y.~S. thanks the sponsorship from Yangyang Development Fund.  SKD acknowledges the support from ANRF, India, under project No. ANRF/ARG/2025/002424/PS. 

\balance
\bibliographystyle{elsarticle-num} 
\bibliography{ref}

\end{document}